\documentclass{article}

\begin{document}

\title{Comments on ``Chen's attractor exists if Lorenz repulsor exists: The Chen system is a special case of the Lorenz system'' [CHAOS 23, 033108 (2013)]}
\author{Guanrong Chen\\\it Department of Electronic Engineering,
\\\it City University of Hong Kong, Hong Kong}
\date{}%  {(Received xx August 2013)}
\maketitle

\bigbreak
{\bf Abstract.}\ This note points out that the assertions of [``Chen's attractor exists if Lorenz repulsor exists: The Chen system is a special case of the Lorenz system,'' CHAOS 23, 033108 (2013)] are groundless and incorrect.
~
\bigbreak
The celebrated general parametric Lorenz system discussed in [1] and herein
is described by
\begin{eqnarray}
\frac{dX}{dt}&=&\sigma (Y-X) \nonumber\\
\frac{dY}{dt}&=&\rho X-Y-XZ \label{eq1} \\
\frac{dZ}{dt}&=&-\beta Z+XY \nonumber
\end{eqnarray}
where $\sigma $, $\rho $ and $\beta $ are real parameters, while the
classical (original) Lorenz system refers to
\begin{eqnarray}
\frac{dX}{dt}&=&10(Y-X) \nonumber\\
\frac{dY}{dt}&=&28X-Y-XZ \label{eq2}\\
\frac{dZ}{dt}&=&-\frac{8}{3}Z+XY \nonumber
\end{eqnarray}
which is chaotic on a small subset\footnote{Due to the structural stability of the Lorenz system, this parameter set can be somewhat enlarged, but using this set does not affect the discussions throughout. The same remark applies to system (\ref{eq4}).} $ \{\sigma ,\rho ,\beta
\}=\{10,28,8/3\}$ inside the 3D real parameter space of the general form
(\ref{eq1}), but for other parameter sets system (\ref{eq1}) may not be chaotic.

The so-called general parametric Chen system discussed in [1,2] and also
herein is described by
\begin{eqnarray}
\frac{dx}{dt}&=&a(y-x) \nonumber\\
\frac{dy}{dt}&=&(c-a)x+cy-xz \label{eq3}\\
\frac{dz}{dt}&=&-bz+xy \nonumber
\end{eqnarray}
where $a$, $b$ and $c$ are real parameters, while the chaotic case is with
the parameter set $\{a,b,c\}=\{35,3,28\}$, giving
\begin{eqnarray}
\frac{dx}{dt}&=&35(y-x) \nonumber\\
\frac{dy}{dt}&=&-7x+28y-xz \label{eq4}\\
\frac{dz}{dt}&=&-3z+xy \nonumber
\end{eqnarray}
Likewise, for other parameter sets of $\{a,b,c\}$, system (\ref{eq3}) may not be
chaotic.

In [1], it is suggested that, with $c\ne 0$, the linear transform
\begin{equation}
\label{eq5}
x=-cX,
\quad
y=-cY,
\quad
z=-cZ,
\quad
\tau =-ct
\end{equation}
which has reversed the time variable if $c>0$, can convert the general Chen
system (\ref{eq3}) to
\begin{eqnarray}
\frac{dX}{d\tau }&=&\tilde {\sigma }(Y-X) \nonumber\\
\frac{dY}{d\tau }&=&\tilde {\rho }X-Y-XZ \label{eq6}\\
\frac{dZ}{d\tau }&=&-\tilde {\beta }Z+XY \nonumber
\end{eqnarray}
where
\begin{equation}
\label{eq7}
\tilde {\sigma }=-\frac{a}{c},
\quad
\tilde {\rho }=\frac{a}{c}-1,
\quad
\tilde {\beta }=-\frac{b}{c}
\end{equation}

By carelessly looking at the algebraic forms of the equations, it is easy to
believe that system (\ref{eq6}) is exactly the same as system (\ref{eq1}), or a special case
of it, as asserted in [1].

However, it must be pointed out that although the parameters set $\{\sigma
,\rho ,\beta \}$ of system (\ref{eq1}) can be the entire 3D real space, for
this nonlinear system its dynamics vary from region to region in the
parameter space, therefore typically it is being investigated case by case
on different parameter sets. Observe, for example, if system (\ref{eq1}) is defined
only on a restricted parameter set with $\beta >0$, which is the case of the
chaotic Lorenz system (\ref{eq2}) where $\beta =8/3$, its third equation
$\frac{dZ}{dt}=-\beta Z+XY$ is stable about zero on the two planes
$X=0$ and $Y=0$, while if system (\ref{eq6}) is defined only on a restricted
parameter set with $\tilde {\beta }<0$, which is the case of the chaotic
Chen system (\ref{eq4}) where $\tilde {\beta }=-3/28$, the corresponding third
equation $\frac{dZ}{d\tau }=-\tilde {\beta }Z+XY$ is unstable about
zero on both planes $X=0$ and $Y=0$. So, even from this simple
observation, one may start to appreciate some subtle differences between the
two systems when they are defined on different parameter sets.

Moreover, it should be pointed out that, in dynamical systems theory,
time reversion is a very critical operation which typically renders a chaotic
system to be non-chaotic. For example, if the following simple form of
transformation (\ref{eq5}) is used on the chaotic Lorenz system (\ref{eq2}):
\[
x=X,
\quad
y=Y,
\quad
z=Z,
\quad
\tau =-t\,,
\]
one obtains a divergent system with all equilibrium stabilities changed to the opposite,
which is by no means equivalent to the original Lorenz system (\ref{eq2}).
This also explains why the concept of ``equivalence'' was never defined through
time reversion in all the chaos theory literature; otherwise, one can be led to 
conclude that ``$\dot x = x$ and $\dot x = - x$ are equivalent'' but obviously their 
stabilities are opposite at zero.

Up to this point, we have only emphasized that although the two general
systems (\ref{eq1}) and (\ref{eq6}) are seemingly identical in their
algebraic forms, their coefficients are defined on different parameter
sets which determine their different dynamics.

Next, observe that although the title, abstract, introduction and conclusion
of [1] all try to create an impression to the readers that the Chen system
is equivalent to (or a special case of) the Lorenz system no matter what, in
the technical text of [1] it is frankly made clear that the two systems (\ref{eq1})
and (\ref{eq3}) are ``equivalent'' (after time reversion) on and only on a particular
parameter set, which is merely a plane in the 3D real parameter space (or a line in
the 2D real parameter plane which [1] discusses):
\begin{equation}
\label{eq8}
\rho +\sigma =-1
\end{equation}
The original statement given in [1] is quoted here for clarity: ``Therefore,
generally, for $c\ne 0$, the Chen system is equivalent to the Lorenz system
in the particular case of the parameter plane $\rho +\sigma =-1$.''

However, obviously many different systems can be equivalent or even identical if they
both are restricted (projected) on a lower-dimensional (parameter) subspace, hence
the above ``equivalence'' does not explain anything meaningful. Even so, an easy catch
is that the chaotic Lorenz system (\ref{eq2}) does not satisfy
this condition at all, where $\rho +\sigma =28+10\ne -1$. With the ignorance of all these, 
it was stated in [1] that ``Thus, for one Lorenz system satisfying $\rho
+\sigma =-1$, there are infinitely many Chen systems'' which satisfy this
condition, including obviously the chaotic Chen system (\ref{eq4}). It is further inferred
in [1] that the general Lorenz system (\ref{eq1}), when its parameters are
restricted to a subset satisfying condition (\ref{eq8}) which does not include the
Lorenz system (\ref{eq2}), is ``equivalent'' to the chaotic Chen system (\ref{eq4}). We do
not follow this logic of inference.

In the study of chaos theory, one is particularly interested in the case
where both systems are chaotic. It has been proved in [3] that the chaotic
Lorenz system (\ref{eq2}) is not smoothly-equivalent to the chaotic Chen system (\ref{eq4}),
namely there does not exist a smooth transform of variables (diffeomorphism) that 
maps one system to another. This statement remains intact unless the proof given in
[3] is found erroneous or one could show a precise counterexample.

To proceed further, in [1] a so-called ``Lorenz repulsor'' is constructed,
as follows: remove the Lorenz chaotic parameter set $\{\sigma ,\rho ,\beta
\}=\{10,28,8/3\}$ from system (\ref{eq1}) or (\ref{eq2}), and replace it by the Chen chaotic
parameter set $\{a,b,c\}=\{35,3,28\}$, then apply transformation (\ref{eq7}) to
result in a ``Lorenz repulsor'' which is in the form of (\ref{eq6}) with $\tilde
{\sigma }=-a/c=-35/28$, $\tilde {\rho }=35/28-1$  and $\tilde {\beta
}=-b/c=-3/28$ (see the Caption of FIG. 2 in [1]). It is clear that this ``Lorenz
repulsor'' is nothing but a linear transformation of the chaotic Chen system
(\ref{eq4}), which is not obtained from the chaotic Lorenz system (\ref{eq2}) per se. In
other words, the assertion of [1] is equivalent to saying that ``if the
attractor of a linearly-transformed Chen system (\ref{eq4}) exists, then the
attractor of the original Chen system (\ref{eq4}) exists''.

Thus, up to this point, according to [1], we are given the following
statements and facts:

\begin{enumerate}
\item The chaotic attractor of the Lorenz system (\ref{eq2}) (with parameters in a small neighborhood of $\{\sigma ,\rho ,\beta \}=\{10,8/3,28\})$ exists by the well-known result of Tucker [4];
\item The parametric Lorenz system (\ref{eq1}) with parameters satisfying $\rho +\sigma =-1$, which does not include the chaotic Lorenz system (\ref{eq2}), is ``equivalent'' to the chaotic Chen system (\ref{eq4});
\item The existence of the attractor of the ``Lorenz repulsor'', precisely a linear transformation of the Chen system (\ref{eq4}) but not of the Lorenz system (\ref{eq2}), implies the existence of the attractor of the Chen system (\ref{eq4}).
\end{enumerate}
Based on the above, it is asserted in (e.g., the Title of) [1]: ``Chen's
attractor exists if Lorenz repulsor exists.'' Obviously, no one is able to
follow this logic of inference. The assertion is groundless and incorrect.

It should be noted that actually we were the first to point out that the
Chen system is a special case of a generalized Lorenz canonical form (see,
e.g., [5] and some other references cited in [1]). In some of our
publications (again, see the references cited in [1]), we already pointed
out many similarities (as well as differences) between the two systems
regarding their dynamics, since they both belong to the same generalized
Lorenz family. If one is satisfied with the knowledge about their
similarities, then there is no need to bother to look at the Chen system.
But for those who have scientific curiosity trying to find out the subtle
differences between the two topologically non-equivalent systems [3],
perhaps one should respect their academic freedom rather than saying that
their works are ``unnecessary or incorrect'' (said in the Abstract and the Lead
Paragraph of [1]).

As a concluding remark, it is not advisable to publish so many similar
commentary articles [7-11] (which are Refs. 56-59 and 63 in [1]), in which
all technical content is the same in nature. Actually, one can (in fact,
should) publish only one commentary and then refer to it in future
publications if necessary. This can relieve the burden of our academic
journals as well.

\end{document}